\documentclass[aps,prb,superscriptaddress,twocolumn,showpacs,amsmath,floatfix,citeautoscript]{revtex4}

\usepackage{graphicx}
\usepackage{float}
\usepackage{dcolumn}
\usepackage{color}
\usepackage{latexsym,bm}
\usepackage[normalem]{ulem}

\begin{document}

\hyphenpenalty=5000

\tolerance=1000

\title{Replica exchange molecular dynamics optimization of tensor network
  states for quantum many-body systems}

\author{Wenyuan Liu}
\affiliation{Key Laboratory of Quantum Information, University of Science and
  Technology of China, Hefei, Anhui, 230026, People's Republic of China}
\affiliation{Synergetic Innovation Center of Quantum Information and Quantum
  Physics, University of Science and Technology of China, Hefei, 230026, China}
\author{Chao Wang}
\affiliation{Key Laboratory of Quantum Information, University of Science and
  Technology of China, Hefei, Anhui, 230026, People's Republic of China}
\affiliation{Synergetic Innovation Center of Quantum Information and Quantum
  Physics, University of Science and Technology of China, Hefei, 230026, China}
\author{Yanbin Li}
\affiliation{Key Laboratory of Quantum Information, University of Science and
  Technology of China, Hefei, Anhui, 230026,  People's Republic of China}
\author{Yuyang Lao}
\affiliation{Key Laboratory of Quantum Information, University of Science and
  Technology of China, Hefei, Anhui, 230026,  People's Republic of China}
\author{Yongjian Han}
\email{smhan@ustc.edu.cn}
\affiliation{Key Laboratory of Quantum Information, University of Science and
  Technology of China, Hefei, Anhui, 230026,  People's Republic of China}
\affiliation{Synergetic Innovation Center of Quantum Information and Quantum
  Physics, University of Science and Technology of China, Hefei, 230026, China}
\author{Guang-Can Guo}
\affiliation{Key Laboratory of Quantum Information, University of Science and
  Technology of China, Hefei, Anhui, 230026,  People's Republic of China}
\affiliation{Synergetic Innovation Center of Quantum Information and Quantum
  Physics, University of Science and Technology of China, Hefei, 230026,
  China}
\author{Lixin He}
\email{helx@ustc.edu.cn}
\affiliation{Key Laboratory of Quantum Information, University of Science and
  Technology of China, Hefei, Anhui, 230026,  People's Republic of China}
\affiliation{Synergetic Innovation Center of Quantum Information and Quantum
  Physics, University of Science and Technology of China, Hefei, 230026, China}
\date{\today }


\begin{abstract}
The tensor network states (TNS) methods combined with 
Monte Carlo (MC) techniques
have been proved a powerful algorithm for simulating quantum many-body
systems. However, because the ground state energy is a highly non-linear
function of the tensors, it is easy to
get stuck in local minima when optimizing the TNS of the
simulated physical systems. To overcome this difficulty,
we introduce a replica-exchange
molecular dynamics optimization algorithm to obtain the TNS ground state,
based on the MC sampling techniques, by mapping the energy function
of the TNS to that of a classical dynamical system. 
The method is expected to effectively avoid local minima.
We make benchmark tests on a 1D Hubbard model based on matrix product states
(MPS) and a Heisenberg $J_1$-$J_2$ model on square lattice based on
string bond states (SBS).
The results show that the optimization method is robust
and efficient compared to the existing results.

\end{abstract}
\maketitle

\date{\today}

\section{Introduction}

Developing efficient methods to simulate strongly correlated quantum many-body
systems is one of the central tasks in modern condensed matter physics.
Recently developed tensor network states (TNS) methods, including the
matrix product states (MPS), \cite{vidal03, vidal04,
fannes92} and the projected entangled pair states (PEPS),
\cite{verstraete04} string bond states(SBS)~\cite{schuch08}, multi-scale entanglement
renormalization ansatz (MERA)~\cite{vidal07}etc. provide a promising scheme to solve
the long standing quantum many-body problems.
In this scheme the variational space can be
represented by polynomially scaled parameters, instead of exponential ones.
Once we have the TNS representation of the many-particle wave functions, the
ground state energies as well as corresponding wave functions can be obtained
variationally.
However, in practice it is still a great challenge to obtain the ground state 
of some complicate physical systems
(such as, frustrated systems and fermionic systems in two dimension)
in the TNS scheme.

Several difficulties reduce the efficiency.
First, though the polynomial scaling, the computational cost respect to the
virtual dimension cut-off $D$ is still very high, particularly, in two and
higher dimensions. For example, the scaling is $D^5$ for periodic MPS
algorithm,\cite{verstraete04_3} $D^{12 - 18}$ for the PEPS algorithm\cite{Murg07}.
For many important systems, in particular, the fermionic systems, the parameter
$D$ should be rather large to capture the key physics. However, limited by
current computation ability, we can only deal with small $D$, typically less
than 10. To overcome this difficulty, in seminal works,
Sandvik et. al\cite{sandvik07} (based on MPS) and Schuch et. al.
\cite{schuch08} (based on SBS) introduced a Monte Carlo (MC) sampling technique
instead of contraction
to reduce the computational cost. The technique
has soon been applied to more general TNS, such as PEPS.~\cite{wang11}
Based on the MC sampling technique, the MPS scaling can be reduced from $D^5$ to $D^3$ for
periodic boundary condition (PBC).~\cite{sandvik07} 
For SBS,\cite{schuch08} the
scaling is O($D^2$) for a 2D open boundary condition (OBC) system and O($D^3$)
for a 2D periodic system with MC sampling technique, which is significantly lower than the standard
contraction methods. For more general PEPS case, the scaling can also be well reduced.~\cite{wang11}

Secondly, the energy function is a highly non-linear function of the tensors.
With the increasing of $D$, the parameter space will be very large and it is very
easy to be trapped in some local minima, when minimizing the energies,
especially for complicated or frustrated systems with a large number of low energy excitations.
Here, we focus on overcoming this difficulty.

In this work, we develop an efficient algorithm to obtain the ground state
energy as well as wave function of a TNS based on the MC sampling method.
We map the quantum many-particle problem to a classical mechanical
problem, in which we treat the tensor elements as the generalized coordinates
of the system. We optimize the energy of the
system using a replica exchange molecular dynamics method.\cite{swendsen86,geyer_book}
By exchanging the system configurations among higher and lower temperatures, it can explore large phase
space and therefore effectively avoid being stuck in
the local minima. The replica exchange method
has been proved very successful in treating classical spin glass~\cite{marinari98}
and frustrated spin systems\cite{cao09} which also suffer from the local minima
problem.  It also has been successfully used to optimize other highly non-linear
problems, such as three-tangle of general mixed states.~\cite{cao10}
Here, we introduce this method in the TNS scheme for the
quantum many-body systems.
We make benchmark tests of the method for a 1D Hubbard model~\cite{fisher89} using
MPS and the 2D $J_1$-$J_2$ Heisenberg model using SBS.~\cite{cirac10}
The results show that this method is efficient and robust. It is also
worth to emphasize that the method introduced here is not limited to the
special type of TNS, but applies to general TNS.\cite{wang11}

\section{Methods}

For simplicity, we describe our method using the example of MPS type of wave
functions. The method can be easily generalized to other types of TNS, e.g.,
SBS ~\cite{schuch08} and PEPS. \cite{wang11}
The many-particle wave functions of one dimensional periodic systems with $N$ sites,
can be written in the MPS\cite{fannes92}, i.e.,
\begin{equation}
  |\Psi_{\rm MPS}\rangle = \sum_{s_1 \cdots s_N=1}^d {\rm Tr} (A_1^{s_1} A_2^{s_2}
  \cdot A_N^{s_N} )|s_1 \cdots s_N \rangle.
\end{equation}
where, $d$ is the dimension of the physical indices $s_k$, and for fixed physical index $s_k$,
$A_k^{s_k}$ are $D\times D$ matrices on site $k$, where $D$ is the Schmidt
cut-off. Given a Hamiltonian $H$ for a system,
the total energy of this system is a function of the tensors at each lattice
site $A_k^{s_k}$, i.e., $E=E(\{A_k^{s_k}\})$.
The main task is to find the ground state wave function and its energy, that is, to find the
global minimum of the function $E(\{A_k^{s_k}\})$ and the corresponding value
of $A_k^{s_k}$.
This problem can be mapped to optimizing the total energy of
a classical mechanical system, in which
the elements of the tensor ${A_k^{s_k}}$ are treated
as the generalized coordinates of the system.
We introduce the Lagrangian of the {\it artificial} system,
\begin{equation}
\mathcal{L}={m \over 2} \sum_{k=1}^N \sum_{s_k=1}^d ||\dot{A}_k^{s_k}||^2-E(\{A_k^{s_k}\}),
\end{equation}
where $m$ is the artificial mass of the ``particles'', and we use $m$=1 in all
the simulations. 
$\dot{A}_k^{s_k}$ is the velocity of corresponding matrix $A_k^{s_k}$
defined on each lattice site.
The norm of matrix $||\dot{A}_k^s||$ is defined as
\begin{equation}
||\dot{A}_k^{s_k}||=\sqrt{\sum_{i,j=1}^D [\dot{a}_{ij}^{s_k}(k)]^* [\dot{a}_{ij}^{s_k}(k)]} ,
\end{equation}
where $\dot{a}_{ij}^{s_k}(k)$ is the velocity corresponding to $a_{ij}^{s_k}(k)$ which is the elements of $A_k^{s_k}$.


We therefore have the Euler-Lagrange equation (we drop the site index $k$ for simplicity),
\begin{equation}
  \frac{d}{dt}\frac{\partial \mathcal{L}}{\partial
    \dot{a}_{ij}^s}-\frac{\partial \mathcal{L}}{\partial {a_{ij}^s}}=0\, ,
\end{equation}
which leads to,
\begin{equation}
m \ddot{a}_{ij}^s = -\frac{\partial E}{\partial a_{ij}^s}.
\label{eq:motion}
\end{equation}
The energy and its derivative respect to given $a_{ij}^s$ can be easily
calculated by MC sampling the physical configuration space.~\cite{sandvik07,schuch08}
Since the MC sampling method for TNS has been described
in details in Refs.~[\onlinecite{sandvik07,schuch08}], we shall not repeat it
here.

Equation~(\ref{eq:motion}) can be solved via the molecular dynamics (MD)
method,~\cite{MD_book} using a velocity Verlet's algorithm,
\begin{equation}
  a_{ij}^s(t+\Delta t)=a_{ij}^s(t)+\frac{1}{2}\Delta t
  [\dot{a}_{ij}^s(t)+\dot{a}_{ij}^s(t+\Delta t)]\, ,
\end{equation}
where,
\begin{equation}
\dot{a}_{ij}^s(t+\Delta t)=\dot{a}_{ij}^s(t)+\frac{1}{2}\Delta t
    [\ddot{a}_{ij}^s(t)+\ddot{a}_{ij}^s(t+\Delta t)]\, ,
\end{equation}
and,
\begin{equation}
m\ddot{a}_{ij}^s(t+\Delta t)=-\frac{\partial E}{\partial a_{ij}^s}(t)\, .
\end{equation}

Now we introduce a temperature $T$ for each tensor $A_k$ as the average
``kinetic'' energies of the ``particles'', i.e.,
\begin{equation}
T=\sum_{i,j=1}^D \sum_{s=1}^d (\dot{a}_{ij}^s)^2/N_D
\label{eq:temperature}
\end{equation}
where $N_D=D^2d$ is the total degree freedoms (number of ``particles'')
of tensor $A_k$. \cite{footnote1}
When the temperature approaches zero, both $\ddot{a}$ and $\dot{a}$
also approach zero, we then obtain the minimum of $E(\{A_k^{s_k}\})$, i.e., the
ground state energy of the quantum system, and corresponding wave function.
When temperature $T$ is sufficiently low, the system can be approximated as
harmonic oscillations around their equilibrium positions, and therefore, 
according to the classical statistics the
total energy of the system is $E(T) \approx E_0+ D^2dT$.

We can run the MD simulation at a given temperature through exchanging energies
with a heat bath. Since we are not interested in the real ``dynamics'' of the
system, one can use the simplest velocity rescaling thermostat:
in order to fix the temperature at $T$, we rescale the velocity
$\dot{a}^s_{ij}$ by a factor $\gamma=\sqrt{T/T^*}$ at each MD step, where
$T^*$ is the instantaneous temperature defined in Eq.~(\ref{eq:temperature}).
Note that when scaling a tensor $A_k$ to $\lambda A_k$, the energy of the
system $E(\{A_k^{s_k})$ remains unchanged. Therefore, we normalize the tensors
by dividing them the largest absolute value of the elements of each tensor 
after each MD step,
to keep the temperature well defined. Furthermore, any change of the tensor $A_k$
that is parallel to $A_k$ during the MD steps have no contribution to the
energy. To improve the efficiency, we orthogonalize the velocity
$\dot{A_k}$ to $A_k$ at each MD step before we rescaling the velocity to the
given temperature,
 \begin{equation}
 \tilde{\dot{A}}_k = \dot{A}_k  - \frac{(\dot{A_k} , A_k)}{(A_k,A_k)} A_k \, ,
 \end{equation}
where the inner product of two matrices is defined as,
\begin{equation}
         (A, B)=\sum_{i,j=1}^D A^*_{ij}  B_{ij} \, .
\end{equation}


Usually the ground state energy of a simple system
can be obtained by a simulated annealing method,
i.e., one starts from a high temperature of the system, and gradually
decreases the temperature to zero. If the temperature cooling is sufficiently
slow, in principle one should get the {\it global} minimum of the system. However, since
the energy is highly non-linear function of the tensors, and for frustrated
physical models, which have many metastable states,
in practice, it often easily be trapped in some local minima.

Here, we adopt the replica exchange (also known as parallel tempering)~\cite{swendsen86, geyer_book} MD
method, which simulates $M$ replicas simultaneously, and each at a different
temperature $\beta_0 = 1/T_{\rm{max}} < \beta_1,\cdots,\beta_{M-2} <
\beta_{M-1}=1/T_{\rm{min}}$
covering a range of interest, 
to avoid being stuck in local minima. Each replica runs independently, except after
certain steps the configuration can be exchanged between neighboring
temperatures, according to the Metropolis criterion,
\begin{equation}
\omega=\left\{
        \begin{array}{ll}
        1 & \Delta H < 0,\\
         e^{- \Delta H} & \rm{otherwise}.
        \end{array}
        \right.
\end{equation}
where $\Delta H = -(\beta_i-\beta_{i-1})(\bar{E}_i-\bar{E}_{i-1})$
in which $\bar{E}_i$ and $\bar{E}_{i-1}$ are the {\it average}
energies of the $i-$th and the $(i-1)$-th replica in a range of certain
MD steps. The inclusion of high-$T$ configurations ensures that
the lower temperature systems can access a broader phase space and avoid being
trapped in local minima.
During the simulation, we keep the  highest temperature $\beta_0$ and lowest
temperature $\beta_{M-1}$ fixed, whereas the rest temperatures distribute
exponentially between the highest and lowest temperatures at the start of
simulation. During the simulation, the temperatures (except $\beta_0$
and $\beta_{M-1}$) are adjusted to ensure that the
exchange rates between the replica are roughly equal.\cite{cao10}

The lower the minimal temperature $T_{\rm{min}}$, the more
accurate the results one can obtain. In principle,
$T_{\rm{min}}$ has to approach zero to get the real ground state. However, decrease
the $T_{\rm{min}}$ will increase the computational cost (the
number of replica temperatures). Instead, one could continue to lower the
temperature sequentially to a desired low temperature,
or adopt a local minimizer (e.g., conjugate gradient method)
after we finish the replica exchange MD simulations, to get more accurate ground state.
 
It is worth noting that a direct use of Monte Carlo method instead of MD
to update the tensors themselves (i.e., one directly change the tensor
elements according to the Metropolis criterion for the total energy)
are not applicable for the scheme.
The reason is that the energy
obtained from MC sampling is not bounded from below.
Therefore, it is very easy to be trapped in a {\it false} energy minimum (i.e.,
the energy minimum due to inadequate MC sampling,
which may be much lower than the real energy of the system),
especially if the sampling is not large enough,
if a Monte Carlo updating method is used.
In contrast, the MD method does not suffer from this problem.

\section{Results and discussion}

In this section, we present the benchmark
tests of our scheme on one-dimensional (1D) Hubbard model and two-dimensional (2D)
$J_1$-$J_2$ model.
Since 1D model has been well studied and has many efficient schemes,
we simply present the results without detailed discussion. 
We discuss more detailed features of the scheme for the 2D model.

\subsection{One-dimensional Hubbard model}

To test our scheme, we compute the ground state energy of the 1D Hubbard model,\cite{hubbard_book}
\begin{equation}
  H = -\sum_{i\sigma} (c_{i\sigma} c^\dagger_{i+1\sigma} +h.c.) + U\sum_{i} n_{i\uparrow}n_{i\downarrow}\,.
\end{equation}
To simulate the 1D Hubbard model, we first transform it to a spin model via Jodran-Wigner transformation.
The many-particle wave function of the ground state of the corresponding spin
model is presented by a MPS.
We optimize the energy using the replica exchange MD method described in
the method section. We use 48 temperatures, with the highest temperature
$T_0$=10$^{-3}$, and lowest temperature $T_{47}$=10$^{-7}$.
The MD time step $\Delta t$=0.5.
During the replica exchange MD, we use 3000$\times L$ MC (each spin flip is
considered a MC sampling)
samplings per MD step, where $L$ is the length of the Hubbard chain.
We further cool down the temperature sequentially
to 10$^{-12}$ to get more accurate ground state energy,
after the replica exchange MD simulation.
The number of MC sampling to calculate the final total energy is 50000$\times L$.

We compare our results to those obtained from the exact diagonalization method in
Table \ref{tab:hubbard}, for a $L$=14 sites, half-filling Fermion Hubbard model, with
PBC, for various $U$ parameters. We take Schmidt cut-off
$D$=6 - 14 for the MPS.
As one can see from the table, we have obtained high accurate
results using the replica-exchange MD optimization method, compared with those
obtained from exact diagonalization method.

\begin{table}[h]
 \caption {The ground state energies calculated by the MD method with various $U$ and
   Schmidt cut-off $D$, compared to the results obtained
  from exact diagonalization.}
	\begin{tabular}{ccccc}
		\hline\hline
		D& U=0.1 & U=1 & U=3 & U=10 \\ \hline
 		4 & -1.25074    & -1.04271 & -0.68991 & -0.26691 \\
 		6 & -1.25728    & -1.04922 & -0.69508 & -0.26845 \\
		8 & -1.25880    & -1.05059 & -0.69608 & -0.26873 \\
 		10 & -1.25903   & -1.05092 & -0.69627 & -0.26875 \\
 		12 & -1.25912   & -1.05100 & -0.69632 & -0.26876 \\
 		14 & -1.25914   & -1.05104 & -0.69633 & -0.26876 \\
 		Exact& -1.25916 & -1.05105 & -0.69634 & -0.26878 \\ \hline\hline
	\end{tabular}
\label{tab:hubbard}	
\end{table}

\subsection{Two-dimensional $J_1$-$J_2$ model}

We simulate the typical two dimensional frustrated spin-$1/2$ Heisenberg model,
namely the $J_1$-$J_2$ model on a square lattice.
The Hamiltonian of the model is,
\begin{equation}
  H = J_1 \sum_{\langle i,j\rangle} {\bf S}_i \cdot {\bf S}_j  + J_2
  \sum_{\langle \langle i,j \rangle \rangle} {\bf S}_i \cdot {\bf S}_j\, .
\end{equation}
The spin operators obey ${\bf S}_i \cdot {\bf S}_i=S(S+1)$=3/4, whereas
$\langle i,j \rangle$ and $\langle \langle i,j \rangle \rangle$
denote the nearest and next-nearest neighbor spin pairs, respectively,
on the square lattice. $J_1$-$J_2$ model has became a promising candidate
model whose ground state may
be a spin liquid state near $J_2/J_1=0.5$.\cite{Wangling1112,wangling13,hongchen12}

Two kinds of generalization to higher dimensions of MPS, i.e., PEPS and SBS can
be used to simulate two dimensional systems.
PEPS has extremely high scaling with the tensor dimension truncation $D$, 
which are $D^{12}$, $(D^{18})$  for OBC
and PBC respectively.\cite{cirac10} 
In contrast, SBS has
much lower scaling to $D$, which are $D^2$ and $D^3$ for OBC and PBC
respectively. Here, as a benchmark, we demonstrate our scheme using the SBS type of wave
functions.

\begin{figure}
 \centering
 \includegraphics[width=3.2in]{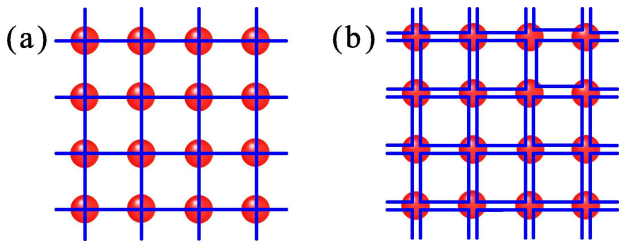}
\caption{The SBS patterns used in the calculations: (a) long strings
and (b) small loops.}
 \label{fig:pattern}
 \end{figure}

The wave functions represented in SBS form can be written as,
\begin{equation}
  | \Psi \rangle = \sum_{s_1 \cdots s_N}^d \prod_{p \in
    \mathcal{P}}{\rm{Tr}}\Big{[} \prod_{x \in p}A_{p,x}^{s_x} \Big{]}
  \Big{|}s_1 \cdots s_N \Big{\rangle},
\end{equation}
where $\mathcal {P}$ is a certain string pattern which contains a set of
strings $p$. The product of matrices $A_{p,x}^{s_x}$ with bond dimension $D$
over $x \in p$ means over the sites $x$ in the order in which appear in the
string $p$. In this work, we use two patterns of the SBS, i.e., the long
strings and small loops as shown in Fig.~\ref{fig:pattern}. 
The two types of SBS satisfy both area law~\cite{schuch08}
and size-consistency.~\cite{wangzhen13}

In our simulations, we use $M$=96 temperatures. Initially, the temperatures
distribute exponentially between the highest ($\beta_0$)
and lowest ($\beta_{M-1}$)  temperatures. For each temperature,
we start from random tensors.
During the simulations, we adjust the temperatures after configuration
exchange for 10 times, whereas there are 300 MD steps between the two
configuration exchanges,
with a step length $\Delta t$=0.01. When sampling
the spin configurations, we enforce $S_z$=0. For each MD step, we sample about 10000
spin configurations. The energies used for temperature exchange
are averaged over 300 MD steps.
We find that adding some small random velocities every 3000 MD steps to
the system can significantly accelerate the convergence,
especially for the large physical systems.
After we finish the replica-exchange MD optimization, we further decrease the
temperature to obtain more accurate results.

 \begin{figure}
 \centering
 \includegraphics[width=2.8in]{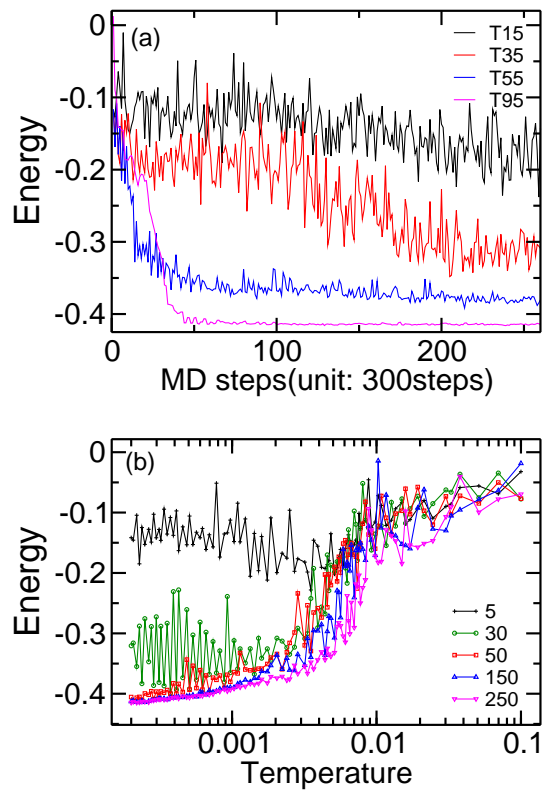}
 \caption { The evolution of the energies of the $J_1$-$J_2$ model with
   $J_2/J_1$=0.7 on a OBC lattice of size 6$\times$6 during MD optimization.
   (a) The energy evolution at different temperatures  as functions of MD step.
   (b) The energies as functions of temperature
     after 5, 30, 50, 150, 250 times of replica exchanges.}
 \label{fig:MD}
 \end{figure}

We simulate the model for $J_2/J_1$$\in$[0, 1], on $N$$\times$$N$ square
lattices, with both OBC and PBC.
To illustrate how the algorithm works, we show in Fig.~\ref{fig:MD}(a),
how the energies evolve during the MD
processes at temperature $T$=$1/\beta_{15}$, $1/\beta_{35}$, $1/\beta_{55}$, $1/\beta_{75}$ and
$1/\beta_{95}$ for a 6$\times$6 OBC system with $J_2$=0.7. We use
with $D$=8 for the long strings, and $D$=4 for the loops.
Note that the exact values
of the temperatures may change during the process.
As expected, the energies of each temperature decreases quickly first to the
``equilibrium'' energy, and then fluctuate around it.
Especially, the energy of the lowest temperature decreases quickly to the energy that near
the ground state energy.
The energy of the system at higher temperatures fluctuate more dramatically than those at
lower temperatures, because the systems have larger ``kinetic'' energy.
By the temperature exchange, it may help the system from being stuck
in some local minima.

Figure~\ref{fig:MD}(b) depicts the total energies as functions
of temperatures after 5, 30, 50, 150, 250 times of temperature exchanges for the
above system.
As we see that the energies at lower
temperatures quickly decrease to near the ground state energy. After enough MD
and temperature exchanges, the energy-temperature curves become
stable. In this situation, we expect that we have obtained reliable ground state.
We then further decrease the temperature to get more accurate ground state energy.

 \begin{figure}
 \centering
 \includegraphics[width=3.0in]{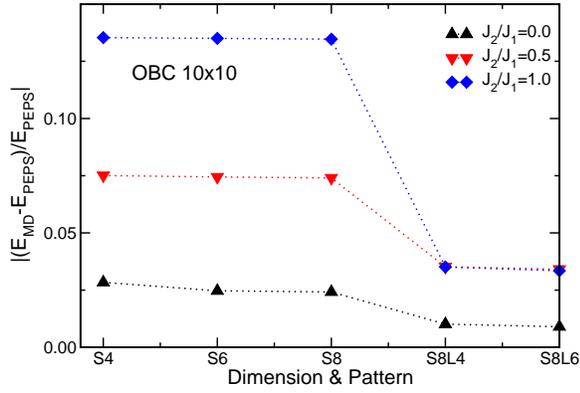}
 \caption{The ground state energies as functions of string patterns
    for the $J_1$-$J_2$ model on a 10$\times$10 OBC lattice. S4 means
   that only long string pattern is used with $D$=4, whereas S8L4 measns that
   both long string pattern and loop pattern are used with $D$=8 for the strings,
   and $D$=4 for the loops.
 }
 \label{fig:convergence}
 \end{figure}

 \begin{figure}
 \centering
 \includegraphics[width=3.0in]{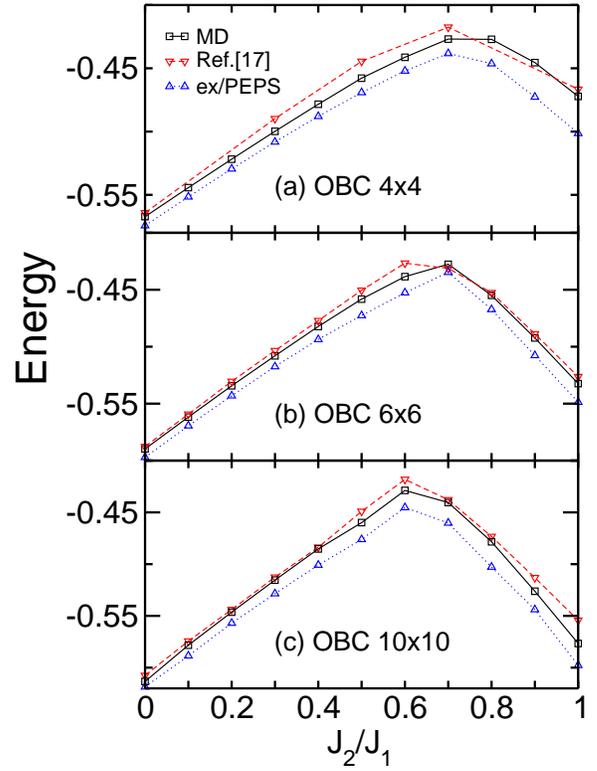}
 \caption{Comparison of the ground state energies of the $J_1$-$J_2$ model on
   the OBC lattices of size (a) 4$\times$4, (b) 6$\times$6, and (c) 10$\times$10.
  The black lines are the results obtained by the replica-MD method, whereas
  the red lines are the results taken from Ref. \onlinecite{cirac10}. The PEPS
 results shown in blue lines for the 6$\times$6, 10$\times$10 lattices are also
 taken from Ref.~\onlinecite{cirac10}.  }
 \label{fig:OBC}
 \end{figure}

 \begin{figure}
 \centering
 \includegraphics[width=3.0in]{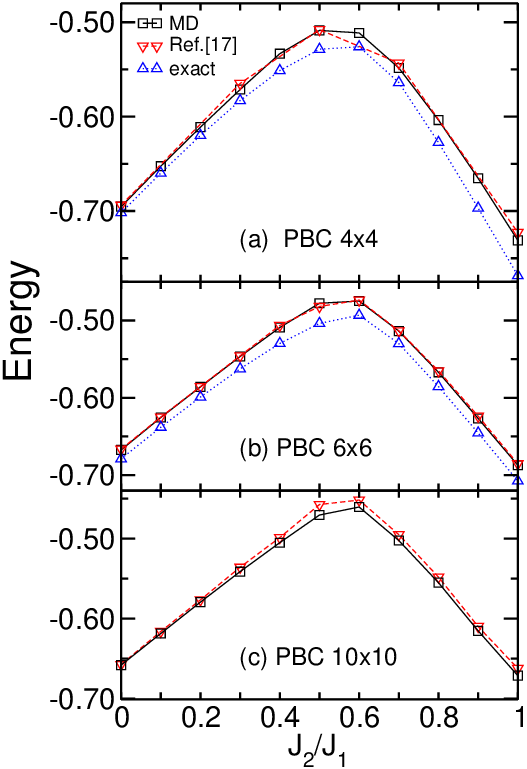}
 \caption{Comparison of the ground state energies of the $J_1$-$J_2$ model for
   the PBC lattices of size (a) 4$\times$4, (b) 6$\times$6, and (c) 10$\times$10.
  The black lines are the results obtained by the replica-MD method, whereas
  the red lines are the results taken from Ref. \onlinecite{cirac10}. The
  exact results of the 4$\times$4, 6$\times$6 lattices are
  taken from Ref.~\onlinecite{schulz94}.  }
 \label{fig:PBC}
 \end{figure}

The improvement of energy by increasing the virtual dimension cut-off $D$ and
adding new SBS patterns are
shown in Fig.~\ref{fig:convergence} for a 10$\times$10 OBC lattice, with
$J_2$=0, 0.5 and 1.0. First, we use only the long strings.
We find that $D$=8 (S8) has converged the results. We then add
the pattern of small loops, and the energies improve significantly.
We find $D$=6 (S8L6) for the loops converge the results.
As one can see that the energy obtained by SBS is still about 1 - 3\% higher
than the exact results or those obtained from PEPS.
This error is from the limitation of the SBS wave functions, and
is not from the optimization process.~\cite{schuch08,cirac10}
Unlike PEPS, the quality of SBS cannot be improved by
simply increasing the dimension $D$ of the tensors alone. However, one can
systematically improve the SBS by adding more patterns of the tensor
strings.~\cite{schuch08} Fortunately, the computational cost increases 
only linearly with the number of string patterns, 
in contrast to the extremely high scaling with the
tensor dimension $D$ in the PEPS method. 
It is very valuable to study how to
improve the SBS wave functions by adding new types of string patterns.
We leave this for future study.

We then compare the obtained results using the method described in the paper
with the results obtained from exact diagonalization method or PEPS
method, and those in Ref.~\onlinecite{cirac10} which also used SBS on lattices of size
4$\times$4, 6$\times$6 and 10$\times$10 with both OBC and PBC.
It can be seen that in all cases the ground state energies optimized by
replica-MD method are improved from the ones obtained using the original optimization
method. For some points,
especially in the strong frustration region, the improvement is significant.

It is worth pointing out that the MD optimization scheme
developed in this work can apply not only to the MPS and SBS types of TNS,
but also to more general TNS,
e.g., PEPS with some modification.~\cite{wang11}
It is suitable to study the systems with rough energy surface, where
other optimization methods may fail.
The current method with MC sampling
has other advantages,  e.g., it is easy to implement the constrains
in physical space.
For example, it is easy to simulate the system in canonical
ensemble using this method, with particle number conservation,
as well as the grand canonical ensemble with fixed chemical potential,
whereas it is difficult to enforce particle number conservation
in the contraction methods.

\section{Summary}

The tensor network states method has been proved a powerful algorithm
for simulating quantum many-body
systems. However, because the ground state energy is a highly non-linear
function of the tensors, it is easy to
be trapped in the local minima when optimizing the TNS of the
simulated physical systems.
We have introduced a replica exchange molecular dynamics method to optimize
the tensor network states.
We demonstrate the method on a one dimensional Hubbard model based on MPS and
the two dimensional frustrated Heisenberg $J_1$-$J_2$ model on square lattices
based on SBS. For the one dimensional model, our results are in excellent agreement
with the those from exact diagonalization method. For the two dimensional model,
our results show improvement over the existing
calculations, especially in the strong frustrated region.
The results demonstrate that our method is efficient and robust.
The method can be generalized to other forms of TNS, e.g. PEPS with some
modification, and provides a useful
tool to investigate complicate many-particle systems, such as
frustrated systems and fermionic systems.

\acknowledgements

LH acknowledges the support from the Chinese National
Fundamental Research Program 2011CB921200, the
National Natural Science Funds for Distinguished Young Scholars
and NSFC11374275.
YH acknowledges the support from the Central Universities WK2470000004,
WK2470000006, WJ2470000007 and NSFC11105135.


\begin{thebibliography}{25}
\bibitem[{foo()}]{footnote1}
\bibinfo{note}{Alternatively, one can define a temperature for the whole system
  in a similar way.}
\end{thebibliography}
\end{document}